 \documentclass[final,5p,times,twocolumn]{elsarticle}

\usepackage{amssymb,amsmath}
\usepackage{lipsum}
\usepackage{xcolor}

\journal{Physics Letters B}

\setcitestyle{number}

\begin{document}

\begin{frontmatter}



\title{Understanding thermalization in a non-Abelian gauge theory in terms of its soft modes}

\author[first]{Sayak Guin\corref{cor1}}
\cortext[cor1]{Corresponding author, email: sayakg@imsc.res.in}
\author[first]{Harshit Pandey}
\author[first]{Sayantan Sharma}
\affiliation[first]{organization={The Institute of Mathematical Sciences, a CI of 
Homi Bhabha National Institute},
            city={Chennai},
            postcode={60113}, 
            country={India}}

\begin{abstract}
We measure the maximal Lyapunov exponent $\lambda_L$ of physical states in a SU(2) gauge theory consisting of soft 
momentum modes both in and out-of-thermal equilibrium conditions 
using ab-initio lattice techniques. We have implemented different algorithms to appropriately describe the 
dynamics of soft-modes for a wide range of temperatures and under non-equilibrium conditions.  The 
non-equilibrium state has been realized starting from an over-occupied initial condition for low momentum soft 
gluons whereas the thermal state comprises of strongly interacting soft gluons at temperatures where these are 
well separated from the hard momentum modes. Spectra of positive Lyapunov exponents is observed in both these states, 
similar to a chaotic dynamical system. From the Kolmogorov-Sinai entropy rate measured in terms of this 
spectrum, we estimate a typical time-scale of $\sim 0.50(3)$ fm/c to achieve thermalization at $T\sim 600$ MeV
starting from the non-thermal state. We also measure, for the first time, the $\lambda_L$ for long wavelength 
critical modes of SU(2) using the out-of-time-ordered correlator of a classical $Z_2$ scalar field 
theory, which shares the same universal behavior with SU(2), near the deconfinement 
phase transition. The $\lambda_L$ is observed to maximize at the transition temperature.

\end{abstract}




\end{frontmatter}




 \section{Motivation}

Understanding the mechanism of thermalization in strongly interacting 
systems is a challenging research problem in theoretical physics~\cite{Berges:2004yj,Berges:2020fwq} 
as it would be able to explain physics at different length scales, e.g. in heavy-ion 
collisions, cold atom traps or in the early universe. Starting from a non-equilibrium 
state, a necessary condition for eventually reaching a thermal state is a mixing between 
different regions of the phase space. A mixing in the phase space can occur if there is 
a mechanical instability in the system~\cite{krylov2014works}, which is characterized by 
at least one positive Lyapunov exponent. Furthermore in the thermal state the entropy or 
the information content~\cite{ShannonPaper} is maximized~\cite{JaynesPaper}. Except for 
a few simple systems~\cite{PhysRevLett.82.520,PhysRevLett.81.1762} it is not very well 
understood how the entropy evolves as a function of time, starting from a chaotic 
non-equilibrium state and eventually maximizing itself in a thermal state.
One important observable is the rate of change of entropy, known as the 
Kolmogorov-Sinai (KS) entropy rate~\cite{KolmogovovEntropy}, which equals the sum 
of all positive Lyapunov exponents for Hamiltonian systems~\cite{pesin2020ljapunov}. 
This equality is known to be true for dissipative systems that have an attractor solution. 
Hence calculating the Lyapunov exponents in chaotic dynamical systems is crucial 
to understand if, and in what time-scales would these eventually thermalize. 

In this Letter, we revisit the question whether non-Abelian gauge theories are  
chaotic dynamical systems~\cite{SAVVIDY1983303,muller1992deterministic,gong1994lyapunov}, 
however, with a goal to understand the phase space mixing and the process of entropy 
maximization in an interacting quantum field theory. For this purpose, we measure the maximal 
Lyapunov exponent of 4-dimensional SU(2) gauge theory using ab-initio lattice techniques 
in three distinct regimes; for a particular non-thermal classical attractor 
state~\cite{Berges:2013eia,Berges:2013fga} and for thermal states near to the 
color-deconfinement phase transition and at high temperatures, deep in the deconfined 
phase. Measuring the KS entropy rate of the soft-gluon modes in a typical 
thermal state at $\sim 600$ MeV from the Lyapunov exponents, we for the first time provide an estimate of 
the typical thermalization time $\sim 0.50(3)$ fm/c for the system to relax into it, starting 
from the classical attractor regime. This estimate is consistent with the value that can be obtained 
using a recently proposed procedure of matching the scales that characterizes the infrared 
magnetic modes of QCD at equilibrium and in a non-thermal state in the self-similar scaling regime~\cite{Pandey:2024goi}. 
Apart from reinforcing the stochastic features of non-Abelian gauge theories~\cite{ebner2024entanglement} 
already strongly developed in its classical modes, our study has a practical relevance as well. Measuring the thermalization time of the chaotic soft-gluon modes thus provides 
an estimate of how long it takes to achieve local thermal equilibration 
i.e. to evolve to a quark gluon plasma phase at temperatures $450$-$600$ MeV, a range which is similar to the 
experimentally measured initial temperature of the fireball formed in a typical heavy-ion collision 
event~\cite{Andronic:2017pug,STAR:2024bpc}, starting from a highly occupied gluon phase.

\section{Theoretical background}

We begin with a discussion of how to measure the (maximal) Lyapunov 
exponent in a non-Abelian SU(2) gauge theory which is a non-linear 
quantum field theory. We recall that, in a classical dynamical system,
starting from two closely spaced initial coordinates and performing 
a Hamiltonian evolution, if the separation between the two trajectories 
increases exponentially with time then the dynamics is defined as chaotic. 
The rate at which the trajectories diverge from each other is characterized 
by the Lyapunov exponent,
$\lambda_L=\lim_{t\rightarrow\infty}\frac{1}{t}\ln\frac{d(t)}{d(0)}$ 
where $d(t)$ denotes the separation between two classical trajectories 
at time $t$. This idea can be easily generalized in a quantum field theory like 
SU(2), in the semi-classical regime, where the gauge links $U_\mu(\mathbf{x})$ 
are analogous to the space coordinates in a classical configuration space. 
Using non-perturbative lattice techniques one can generate ensembles of gauge 
links using Monte-Carlo algorithms. In thermal equilibrium, two different gauge 
configurations which are separated by infinitesimal Monte-Carlo time steps can be 
evolved in real-time using the SU(2) Hamiltonian in order to estimate the Lyapunov 
exponent. 

Alternately, another definition of the Lyapunov exponent can also be motivated by studying the following 
out-of-time ordered correlator (OTOC) defined in one-dimensional classical 
dynamics as, 
\begin{equation}
\{x(t),p(0)\}^2=\left(\frac{\delta x(t)}{\delta x(0)}\right)^2\sim 
\rm{e}^{2\lambda_L t}~,
\label{eqn:otocqm}
\end{equation}
The OTOCs can be easily generalized for quantum field theories. For real-valued 
scalar fields, the expression in the LHS of Eq.~\ref{eqn:otocqm} can be re-written by replacing the Poisson bracket in terms of 
position and momenta with a commutator of the field operator 
$\hat{\phi} (\mathbf{x},t)$ and their conjugate momentum fields 
$\hat{\pi} (\mathbf{x},t)$. However for scalar fields the OTOC can also be equivalently defined as 
$[\hat\phi(\mathbf{x},t), \hat \phi(\mathbf{0},0)]^2$~\cite{Schuckert:2019oao}. 
In a quantum system, OTOC grows exponentially upto some time $t_E$, called the 
Ehrenfest time~\cite{rammensee2018many} beyond which its growth saturates. This 
is the timescale during which the wavefunction has spread over the entire volume of 
the quantum system and there is a transition from a wave-like to a particle-like 
behavior~\cite{hashimoto2017out}. However in the semi-classical 
limit, the OTOC will not saturate. Such a scenario is realized when scalar fields have 
large occupation numbers, where one can replace the commutator  with the following expression,
\begin{equation} 
 \label{eqn:otocdef}
    C(\mathbf{x},t)=\langle\{\phi(\mathbf{x},t),
    \phi(\mathbf{0},0)\}^2\rangle
    =\langle \left(\frac{\delta\phi(\mathbf{x},t)}{\delta
    \pi(\mathbf{0},0)}\right)^2 \rangle~.
   \end{equation}
We henceforth denote the classical fields $\phi(\mathbf{x},t)$ as $\phi_\mathbf{x}(t)$ 
and similarly for its conjugate fields. If indeed the system is chaotic, the 
(maximal) Lyapunov exponent $\lambda_L$ can be extracted from the OTOC using 
the relation,
\begin{equation}
    \lambda_L=\lim_{t\rightarrow\infty}\frac{1}{2t}\ln~ 
C(\mathbf{x=0},t)~.
\label{eqn:lyapunov}
\end{equation}
An OTOC typically spreads ballistically within the light cone along rays of constant 
speed $\mathrm{v}=x/t$ in the $x$-$t$ plane with a scaling exponent $\nu$~\cite{khemani2018} as,
\begin{equation} 
    C(\mathbf{x},t)  \sim \rm{e}^{2\lambda(v)t}~,~\lambda(v)=\lambda_L ~\left[1-\left(\frac{v}{v_B}\right)^\nu\right]~.
\label{eqn:lightconespread}
\end{equation}
From the spectrum of the Lyapunov exponents $\lambda(\mathrm{v})$ in the velocity space of propagating 
OTOCs, one can define the butterfly velocity $\mathrm{v}_B$ from the relation $\lambda(\mathrm{v}_B)=0$. 
The quantities $\lambda_L, \mathrm{v}_B$ and exponent $\nu$  are all, in general, temperature 
dependent. The diffusion coefficient $D$ is related to the maximal Lyapunov exponent 
$\lambda_L$ and the butterfly velocity $\mathrm{v}_B$ through the relation~\cite{blake2016universal},
\begin{equation} \label{diff}
    D=\frac{\mathrm{v}_B^2}{\lambda_L}~.
\end{equation}
In Section~\ref{sec:results}, we will extract $\lambda_L, \mathrm{v}_B$ and $D$ for SU(2) 
gauge theory in thermal equilibrium for a wide range of temperatures as well as in 
a particular non-equilibrium state. The ab-initio lattice algorithms used for this 
purpose are detailed in the next section.

\section{The algorithm }

\subsection{Studying highly occupied SU(2) gauge theory under non-equilibrium conditions}
\label{sec:SU2noneq}
For a specific realization of a non-equilibrium state of SU(2) gauge theory, we start from 
an initial condition where modes of gauge fields the momentum $\mathbf{p} $
are occupied according to a phase-space distribution given by, 
$\tilde f(\mathbf{p})= g^2 f(\mathbf{p})=n_0\frac{Q_s}{\vert \mathbf{p} \vert}\rm{e}^{\frac{-\vert \mathbf{p} \vert^2}
{2Q_s^2}}$ ~\cite{Pandey:2024goi}. Here $Q_s$ is the gluon-saturation scale which is typically between 
$1$-$2$ GeV and $g$ is the gauge coupling. We consider $Q_s=1.5$ GeV in this work. The choice 
of the initial phase-space distribution becomes irrelevant after a time scale $Q_st\sim n_0^{-2}$, which corresponds 
parametrically to the inverse of the large angle scattering rate between the quasiparticles~\cite{Mace:2016svc,Schlichting:2012es,Berges:2013fga}.  Here and in the subsequent 
sections we would be considering the theory to be discretized on a 3-dimensional spatial box with $N$ sites 
along each spatial direction and $a$ as the lattice spacing. The quantity $n_0$ sets the initial occupation 
number of the infrared modes of the gauge fields which we have chosen such that it is close to the typical 
equilibrium energy densities at high temperatures $T\gtrsim 600$ MeV. Since the occupation numbers are 
large (owing to g being small), the gauge links behave classically. For generating these classical configurations, 
we solve for the Hamilton's equations of motion for the gauge links $U^{i}_\mathbf{x}(t)$ and its conjugate fields 
$E^{ib}_\mathbf{x}(t)$  on the lattice along directions $i=1,2,3$ and with color components $b=1,2$,
\begin{eqnarray}
\nonumber
    E^{ib}_{\mathbf{x}}\left(t+\frac{\Delta t}{2}\right)&=&
    E^{ib}_{\mathbf{x}}\left(t-\frac{\Delta t}{2}\right)
    +2\frac{\Delta t}{a}\sum_{j\neq i}^{N^3} \text{tr}\left[i\Gamma^{b} (V^{ij}_{\mathbf{x}}+c.c)\right]~,\\
           U^i_\mathbf{x}(t+\Delta t)&=&
           \rm{e}^{i\frac{\Delta t}{a}~E^{ib}_{\mathbf{x}}\left(t+\frac{\Delta t}{2}\right)\Gamma^b} U^i_\mathbf{x}(t)~,
    \label{eqn:classicalEevol2}
\end{eqnarray}
where $\Gamma^b$ are generators of SU(2) and $V^{ij}_{\mathbf{x}}$ is the spatial staple defined in the $i$-$j$ 
plane as the derivative of the gauge plaquette $U_P$, defined using path ordering $\mathcal{P}$ in color space as,
\begin{equation}
    V^{ij}_{\mathbf{x}}=\frac{\delta U_P}{\delta U^{i}_{\mathbf{x}}}~,~
    U_P^{ij}=\mathcal{P} \left(U^{i}_{\mathbf{x}}U^{j}_\mathbf{x+\hat{i}}U^{i\dag}_\mathbf{x+\hat{j}}U^{j \dag}_{\mathbf{x}}\right).
\end{equation}
The initial conditions for the gauge and electric fields are chosen as a
classical-statistical ensemble representing vacuum fluctuations of transversely polarized 
gluons~\cite{Mace:2016svc}. We use a symplectic leap-frog integrator for the time evolution 
in order to respect time-reversal invariance, with a time-step $\Delta t=0.0125 Q_s$. We study 
different lattice volumes with extent along the spatial directions $N=128$, with  spacings 
$aQ_s=1.115, 0.9$ such that corresponding effective temperatures match with the thermal 
states at temperature $T$ which were generated using the algorithm discussed in the next section i.e. 
$ T_{\mathrm{eff}} \approx T$. The spatial length of the box in physical units is $\sim 16$ fm. 
The Gauss law  $ \frac{a}{\Delta t}\sum_{i}\text{tr}
i\Gamma^{a}\left[ U_{\mathbf{x}}^i(t)-U^i_{\mathbf{x}+\hat{i}}(t)\right]$ was implemented with a 
precession of $10^{-15}$ at $t=0$ and it was checked to remain so at later times. With such a choice of 
initial condition the system eventually enters a self-similar scaling regime at late times where 
the gluon distribution function takes the form $\tilde f(\mathbf{p},t)=(Q_st)^{-4/7}f_s((Q_st)^{-1/7}\mathbf{p})$, 
where $f_s$ is a homogeneous function describing a static distribution~\cite{Berges:2013fga}. 
Within such a regime the hard, electric and the magnetic momentum scales, the latter two 
corresponding to scales set by the Debye mass and the square root of the spatial string tension respectively, 
are well separated similar to a thermal plasma~\cite{Berges:2023sbs}.

In order to measure the degree of chaoticity in such a system in a gauge-invariant manner, we follow 
the procedure as outlined in Ref.~\cite{Pandey:2024goi}. First we implement an infinitesimal shift 
in the initial condition to $n_0+\delta n_0$ where $\delta n_0=10^{-5}$ and perform a Hamiltonian 
evolution of the gauge links $U^{\prime}_\mathbf{x}(t)$ as a function of time. We next measure 
the separation between two gauge trajectories at a time $t$ starting with these two 
infinitesimally close initial conditions, through the gauge-invariant distance measure defined 
in terms of the spatial plaquettes,
\begin{equation}
    d(t)=\frac{1}{2N_p}\sum_P \vert \text{tr} U_P(t)- \text{tr} U^{\prime}_P (t)\vert \sim e^{\lambda_Lt}
\end{equation} 
where $N_P$ are the number of independent plaquettes. We particularly choose the time interval to be 
within the self-similar scaling regime, in order to extract the Lyapunov exponent.

\subsection{Studying thermal SU(2) plasma at high temperatures}
At finite temperatures, apart from the hard scale $\pi T$ present naturally in the problem, 
there are two new scales that are generated dynamically: the electric scale $gT$ 
and the magnetic scale $g^2 T/\pi$~\cite{Laine:2016hma}. When these three scales are 
well-separated, one can integrate out the hard and electric gluons with momenta $\gtrsim gT$ 
to obtain an effective Hamiltonian~\cite{Bodeker:1998hm}, which describes 
the dynamics of the magnetic (soft) gluons with the color electric fields 
$\tilde E_{\mathbf{x}}^{i}$ satisfying the equation of motion 
\begin{equation}
-\partial_t \tilde E_{\mathbf{x}}^{i}+[D_j,\tilde F^{ji}(\mathbf{x})]=\sigma \tilde E^{i}_{\mathbf{x}}+\tilde \zeta^{i}_{\mathbf{x}}(t)~,
\label{eqn:langevinevogauge}
\end{equation}
where $\sigma$ is the color conductivity and $\tilde \zeta^{i}_{\mathbf{x}}(t)$ are the stochastic 
color-force fields which satisfy $\langle\tilde\zeta^{ic}_{\mathbf{x_1}}(t_1)\tilde\zeta^{jb}_{\mathbf{x_2}}
(t_2)\rangle=2T\sigma\delta^{ij}\delta^{cb}\delta^{3}(\mathbf{x_1-x_2})\delta(t_1-t_2)$
in accordance to the fluctuation-dissipation relation. Discretizing the Eq.~\ref{eqn:langevinevogauge}, 
the electric fields, noise fields and color-conductivity can be written in dimensionless units  
on the lattice in terms of $E_{\mathbf{x}}^{i}=g \tilde E_{\mathbf{x}}^{i}a^2,~
\zeta^{i}_\mathbf{x}=a^3 g\tilde\zeta^{i}_\mathbf{x}(t)$ and $\sigma a=\frac{\sigma}{T}.T a$ respectively. 
We set $\sigma/T$ to its perturbative values~\cite{Arnold:1999uy}, given by the relation
\begin{equation}
\sigma^{-1}=\frac{3N_cg^2T}{4\pi m_D^2}\left[\ln \frac{m_D}{\gamma}+3.041\right]~,
\label{eqn:colorconductivityDef}
\end{equation}
where the coupling $g$ 
at each temperature $T$ is determined using the 2-loop $\beta$ function and $1/\gamma$ is the mean free path of the color changing processes which varies between $\gamma/m_D\in [0.883, 1.034]$
for the range of temperatures considered in our work. Here $N_c=2$ and the Debye mass is defined at one-loop 
order in perturbation theory i.e., $m_D=\sqrt{\frac{2}{3}}gT$. 
We numerically implement Eq.~\ref{eqn:langevinevogauge} in dimensionless units which is 
similar to the electric field evolution in Eq.~\ref{eqn:classicalEevol2} but with additional contributions due to the 
noise and damping terms given by 
$- \Delta t/a\left[a\sigma E^{ib}_{\mathbf{x}}(t-\Delta t)-\mathbf{\zeta}^{ib}_{\mathbf{x}} (t-\Delta t)\right]$. 
The evolution in Eq.~\ref{eqn:langevinevogauge} is similar to a Langevin equation for magnetic 
modes of the gauge fields in presence of a heat bath. Here, the noise term represents the effects of the hard 
modes which form the heat bath and the damping dynamically brings the system to a thermal equilibrium. This  
classical system consists of $6N^3$ oscillators of energy $a T$ and has an energy density on the lattice~\cite{gong1994lyapunov} given by 
$\frac{6}{N^3}\sum_{\mathbf{k}}|\mathbf{k}|\frac{a T}{|\mathbf{k}|}~,$
where the sum is over all allowed lattice momenta $\mathbf{k}$, which we have also verified in our calculations.
On the other hand, energy density of the quantum SU(2) gauge theory at temperatures $\sim 2$ times the 
deconfinement temperature is close to its Stefan-Boltzmann limit $\pi^2T^4/5$~\cite{Fingberg:1992ju}. The lattice 
spacing in the effective theory is set in physical units using the criterion that the measured energy density  
matches with the quantum theory at each $T$. This results in a condition $T.a=(30/\pi^2)^{1/3}$, which we 
use to set the lattice spacing in physical units. Once the gauge links are thermalized at a temperature $T$, there 
is an onset of a plateau in the values of the plaquette as a function of Langevin time. We then consider two 
thermal configurations infinitesimally separated as a function of Langevin time, switch off the noise and the 
damping terms in the Hamiltonian and evolve them subsequently as a function of real-time.  This is to implement 
non-equilibrium evolution of the system of soft gluons along the Schwinger-Keldysh contour, which for this case,
consists of only forward-in-time propagation since soft gluons have large occupation numbers. We then calculate 
the gauge invariant distance measure $d(t)$ introduced earlier to calculate the (maximal) Lyapunov 
exponent.

\subsection{Chaotic dynamics near deconfinement phase transition}

Near the deconfinement phase transition temperature $T_c$ the hard, electric and magnetic 
scales coincide, hence the effective thermal field theory described in the preceding section 
can no longer describe the gauge field dynamics. Hence we change our strategy to estimate the  
Lyapunov exponent. Since SU(2) gauge theory shares the same universality class as the 3D Ising 
model~\cite{Svetitsky:1982gs}, hence its long-wavelength or low energy critical modes will 
undergo the same dynamics as a $Z_2$ scalar field. We thus simulate a relativistic $Z_2$ scalar 
field theory to understand the chaotic dynamics of the critical modes near the deconfinement phase 
transition. We start with the Hamiltonian defined on the lattice as,
\begin{equation} \label{eqn:Z2Hamiltonian}
    H=\sum_\mathbf{x} \left[\frac{\pi_\mathbf{x}^2}{2}-\frac{1}{2}
    \sum_\mathbf{y\sim x}^{} \phi_\mathbf{x}\phi_\mathbf{y}+\left(\frac{m^2 a^2}
    {2}+3\right)\phi_\mathbf{x}^2+\frac{\lambda_s}
    {4!}\phi_\mathbf{x}^4\right]~.
\end{equation}
Here $\sum_\mathbf{y\sim x}$ implies the sum over all nearest neighbors $\mathbf{y}$ around the lattice 
site $\mathbf{x}$. In our work, we set the values of $m^2a^2=-1~,~\lambda_s=1$.  We first generate thermal 
ensembles using the Langevin dynamics, in which we consider the time evolution of the scalar and its conjugate 
momentum fields through equations of motion,
\begin{eqnarray}
  \partial_{t} \phi_\mathbf{x} (t) &=& \frac{\partial H}{\partial \pi_\mathbf{x}}~=~\pi_\mathbf{x}(t)~,\\
 \partial_{t} \pi_\mathbf{x} (t) &=&-\frac{\partial H}{\partial \phi_\mathbf{x}}-
\gamma\pi_\mathbf{x}(t)+\eta_\mathbf{x}(t)~\sqrt{2\gamma T}~.
\label{eqn:langevinevolscalar1}
\end{eqnarray}
where $\gamma$ is the damping term and $\eta_\mathbf{x}$ are Gaussian noise fields defined 
on each lattice site with a zero mean value and satisfying
$\langle \eta_\mathbf{x}(t)\eta_\mathbf{y}(t^\prime)\rangle=\delta_\mathbf{xy} \delta(t-t^\prime)$.
The relation between the damping factor $\gamma$ and the coefficient of the 
noise term mimicking thermal fluctuations can be obtained using the fluctuation-dissipation 
theorem. To initialize the algorithm, we select random initial values of $\phi_\mathbf{x}(0)$ and 
 $\pi_\mathbf{x}(0)$ on each lattice site $\mathbf{x}$. We then numerically evolve the above 
 differential equations using the symplectic leap-frog integrator with 
 a time step $\Delta t=0.01$.  Since Langevin dynamics is ergodic and 
 follows the condition of detailed balance, we expect 
 the system to thermalize after some Langevin time $t_L$. Once we achieve thermalization 
 using the Langevin dynamics, we take the thermal configurations of the scalar fields and perform a real-time 
classical evolution. Further, we implement
infinitesimal perturbation about the initial thermal configurations, denoted 
by $\delta\phi_\mathbf{x}(0)=0$ and $\delta \pi_\mathbf{x}(0)=r\delta_{\mathbf{x0}}$, 
where $r$ is a Gaussian random number between $0$ and $0.01$. Since 
$\delta \pi_\mathbf{x}(t)=\partial_t \delta \phi_\mathbf{x}(t)$
one can re-express the initial momentum profile in the configuration space 
at $\mathbf{x}\equiv (0,0,0)$ and $t=0$ and then evolve the fluctuations of the scalar field via
\begin{equation} 
    \partial_t^2 \delta\phi_\mathbf{x}(t)=\nabla^2\delta\phi_\mathbf{x}(t)-
     m^2\delta \phi_\mathbf{x}(t)-\frac{\lambda_s}{2}\phi^2_\mathbf{x}(t)
     ~\delta\phi_\mathbf{x}(t)~.
    \label{eqn:deltaphievol}
\end{equation}
In Eq.~\ref{eqn:deltaphievol} one requires the values of the scalar 
fields $\phi_\mathbf{x}(t)$ as an input which are obtained through a 
classical Hamiltonian evolution of initial thermalized values of 
$\phi_\mathbf{x}(0)$ at each temperature $T$. We further evolve the 
initial momentum field fluctuations 
$\delta\pi_\mathbf{x}(t)$ using $\partial_t\delta \phi_\mathbf{x}(t)=\delta \pi_\mathbf{x}(t)$. 
We thus have all the ingredients to calculate OTOC in the semiclassical limit using 
Eq.~\ref{eqn:otocdef}, from which we extract the Lyapunov exponent, as discussed in the 
next section.

\begin{figure}

    \centering
    \includegraphics[width=0.48\textwidth]{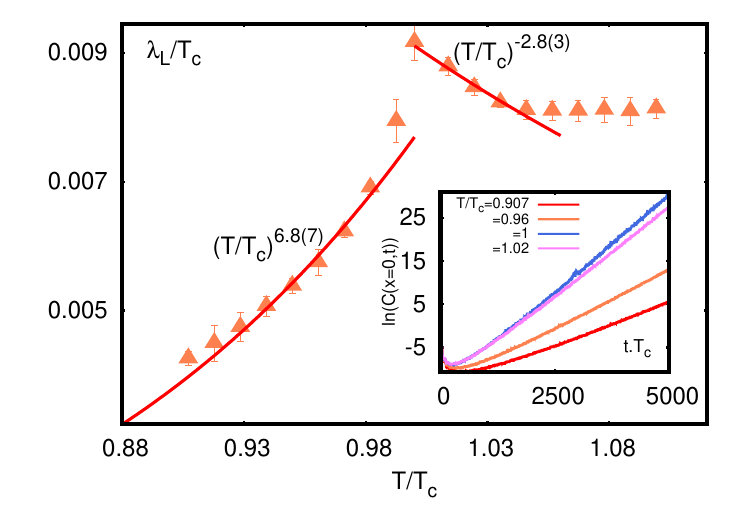}
    \caption{The Lyapunov exponent as a function of $T/T_c$ for a $Z_2$ scalar field 
    theory, which from universality represent the critical long-distance modes of a $SU(2)$ 
    thermal state near the deconfinement transition temperature $T_c$. The inset shows the growth of 
    OTOC with time $t.T_c$.}
    \label{fig:LyaScalar}
\end{figure}

\section{Results}
\label{sec:results}

We first measure the OTOCs for $Z_2$ scalar field theory at different 
temperatures close to the phase transition, shown in the inset of 
Fig~\ref{fig:LyaScalar}. After an initial dip, the OTOCs exhibit an exponential 
rise as a function of time. By performing a fit to the OTOCs at late times 
$t.T_c>1000$ with an exponential ansatz, we extract the (maximal) Lyapunov exponent 
at different temperatures, results of which are shown in Figure~\ref{fig:LyaScalar}. 
The Lyapunov exponent thus extracted varies as $\lambda_L\sim T^{6.8(7)}$ below $T_c$, 
whereas in the range $(1,1.05) T_c$ its temperature variation can as characterized 
as $T^{-2.8(3)}$, which was determined after performing a fit to our data. In our 
calculations $a.T_c=9.37$ which agrees very well with the values obtained in the 
literature~\cite{schweitzer2020spectral}. For reference, the deconfinement phase 
transition temperature in SU(2) is $T_c\sim 300$ MeV~\cite{Fingberg:1992ju}.

\begin{figure}
    \centering
    \includegraphics[width=0.46\textwidth]{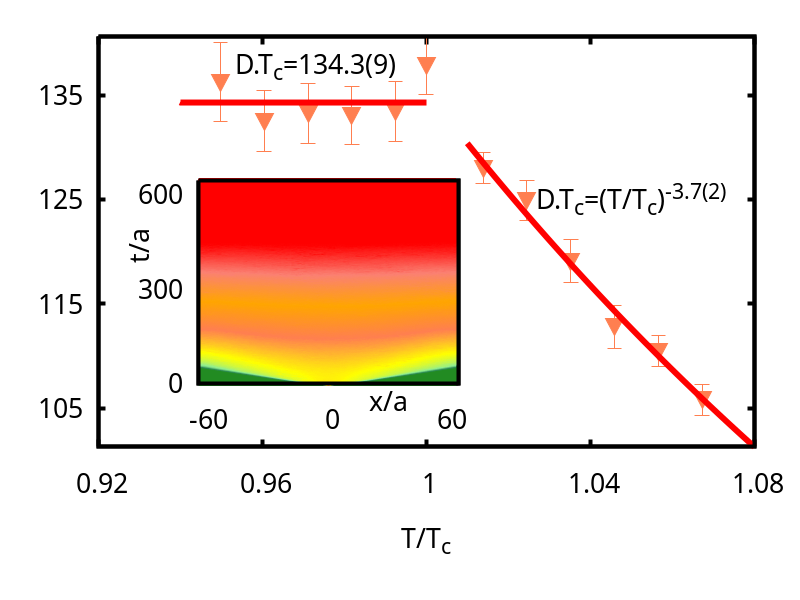}
    \caption{The temperature dependence of the diffusion coefficient 
    $D$ for $Z_2$ scalar fields near $T_c$, extracted using Eq.~\ref{eqn:lightconespread}. The inset 
    shows the ballistic spread of the OTOC of $Z_2$ scalar fields 
    in $x$-$t$ plane at $T_c$. The color profile from green to red denotes 
    increasing values of OTOC in the $x$-$t$ plane.}
    \label{fig:DiffusionCoeff}
\end{figure}

Further, in the inset of Fig~\ref{fig:DiffusionCoeff}, we show the heat-map of the 
OTOC at $T=T_c$ in a 2-dimensional plane as a function of time $t$ and one of the 
spatial coordinates, performing a sum over the other spatial coordinates.
The OTOC spreads ballistically within the light cone from an initial small 
perturbation at the origin as evident from the color gradient of the heat-map. 
In general, $\lambda_L$ is a function of the velocity and it can be 
measured from the OTOC using Eq.~\ref{eqn:lightconespread}. 
As evident in Fig~\ref{fig:LyaScalar}, the Lyapunov exponent is maximum 
along $\mathrm{v}=0$ i.e. along the time direction where the rate of increase in 
the velocity is the largest compared to any other direction. We next calculate the butterfly 
velocity $\mathrm{v}_B$ by fitting the OTOC $C(\mathbf{x},t)$ to the ansatz given in 
Eq.~\ref{eqn:lightconespread}.  The $\mathrm{v}_B$ varies across the phase 
transition and is found to be maximum at $T=T_c$. In Fig~\ref{fig:DiffusionCoeff}, 
we have shown the diffusion coefficient as a function of temperature, which is 
related to the butterfly velocity through the relation in Eq.~\ref{diff}. 
As evident from the figure, the diffusion coefficient is constant below $T_c$ but decreases 
with increasing temperatures as $(T/T_c)^{-3.7(2)}$ for $T>T_c$.

\begin{figure}
    \centering
    \includegraphics[width=0.48\textwidth]{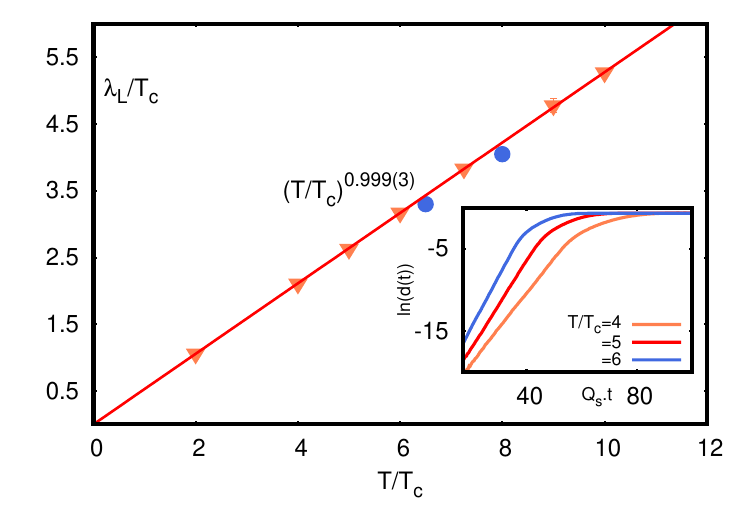}
    \caption{The Lyapunov exponent as a function of $T/T_c$ for a SU(2) gauge  
    theory in thermal equilibrium at high temperatures $T>>T_c$, shown as triangles. These 
    are compared with the Lyapunov exponents in a non-thermal state of SU(2) with 
    comparable energy densities (circles). The inset shows the growth of the gauge invariant 
    distance measure with time $t.T_c$ for a thermal state of SU(2) for 3 different 
    temperatures.}
    \label{fig:lyapunov_SU2}
\end{figure}

At higher temperatures, $T \gg T_c$, the gauge invariant distance measure $d(t)$ was
measured on a lattice with $N=32$ spatial sites and for different temperatures, results 
for which are shown in the inset of Fig~\ref{fig:lyapunov_SU2}. The initial $d(t)\sim 10^{-15}$ 
measures the typical difference between the plaquettes in two thermal configurations 
which, when evolved in time, grows exponentially eventually saturating  
at late times to $\sim 10^{-2}$, since the trace of plaquette is bounded to unity 
and the gauge group is compact. Within this timescale, the 
momentum mode occupancies of gluons remain similarly close to the thermal distribution. 
The lattice size is chosen to be large enough such that the 
physical lengths are $\sim 15(6)$ fm at temperatures $0.6(1.5)$ GeV respectively. We extract 
the (maximal) Lyapunov exponent $\lambda_L$ from the exponentially rising part of 
$d(t)$ and study its temperature dependence, results of which are shown in 
Fig.~\ref{fig:lyapunov_SU2}. The $\lambda_L$ varies linearly with temperature in the range
$0.6$-$3$ GeV with a slope $\lambda_L/T \sim 0.52$ which satisfies the conjectured bound 
$\lambda_L\leq 2\pi T$ in quantum many-body chaotic systems~\cite{maldacena2016bound} and 
also satisfies $\lambda_L \sim g^2 \mathcal{E}/6$, where $\mathcal{E}$ is the average energy 
per plaquette~\cite{muller1992deterministic}. The diffusion coefficient 
extracted from $\lambda_L$ using Eq.~\ref{diff} varies inversely with temperature as $(T/T_c)^{-0.95(3)}$ shown 
in Fig~ \ref{fig:diffusion_SU2}. At these temperatures the distance measure spreads from an initial 
tiny fluctuation at $t.T=0$ spreads ballistically over the space-time and grows exponentially within the light-cone, 
evident from the inset of the same figure, the color gradient from green to red indicates increasing values 
of the distance measure $d(t)$.  Across these high temperatures, the butterfly velocity is $\mathrm{v}_B=0.8c$ 
and does not vary significantly. Compared to its value at $T_c$, the butterfly velocity reduces by only $\sim 10\%$ 
which implies that once thermalized, most of the configuration (color) space gets occupied.

We next extract the Lyapunov exponent within the self-similar scaling regime of the non-equilibrium 
SU(2) plasma as outlined in Sec.~\ref{sec:SU2noneq}. In order to compare with a typical thermal state 
at temperature $T\sim 2$ GeV, we set the initial density of gluons to be $n_0=16$, and consider lattice 
box of spatial extent $\sim 16$ fm such that the energy densities are similar in the thermal as well in 
the non-thermal scaling regime. The (maximal) Lyapunov exponent $\lambda_L=0.66~Q_s$ extracted in this non-thermal 
state is similar in magnitude to that in a thermal state at $T\sim 2$ GeV. We have also shown the values of the 
maximal Lyapunov exponents in Fig~\ref{fig:lyapunov_SU2} in this non-thermal state, as a function of 
the effective temperature extracted from the fourth root of the energy density, as blue data points. The data 
points lie on the same curve that characterizes the temperature dependence of the thermal Lyapunov exponents. 
For lower energy densities $\varepsilon(\sim \sqrt{n_0} \lesssim 1)$ it was earlier 
conjectured~\cite{chirikov1981stochastic} and later observed~\cite{Pandey:2024goi} from 
classical-statistical lattice calculations in the self-similar non-thermal scaling regime of SU(3), that 
the (maximal) Lyapunov exponent $\lambda_L$ varies as $\varepsilon^{1/4}$. Our calculations performed 
in the non-thermal scaling regime of SU(2), with a larger initial gluon energy density $(\sim \sqrt{n_0} \gtrsim 4)$ 
also yield a linear scaling of $\lambda_L$ with $\varepsilon^{1/4}$. The linear scaling of $\lambda_L$ at higher 
energies similar to a thermal plasma can be understood in terms of linear temperature dependence of the 
plasmon damping rate~\cite{biro1995lyapunov}.

\section{Physical implications of our results}

Our results demonstrate that the infrared (soft) modes of SU(2) gauge theory exhibit chaotic dynamics both in 
thermal equilibrium at a wide range of temperatures $0.9$-$10~T_c$ as well in a particular non-thermal fixed 
point as evident from the positive (maximal) Lyapunov exponent $\lambda_L$. Interestingly the $\lambda_L$ 
decreases beyond the critical temperature $T_c$ with temperatures $T>T_c$, maximizing at the deconfinement phase transition. 
Such a property of $\lambda_L$ for the critical modes has been observed earlier in spin systems~
\cite{bilitewski2018temperature,bilitewski2021classical,ruidas2021many,butera1987phase} as well as in scalar 
field theories in 2 dimensions~\cite{Schuckert:2019oao}. The temperature dependence of $\lambda_L$, thus, 
is not particularly related to the details of microscopic interactions between the degrees of freedom but 
to the symmetries of the system.

\begin{figure}
    \centering
    \includegraphics[width=0.48\textwidth]{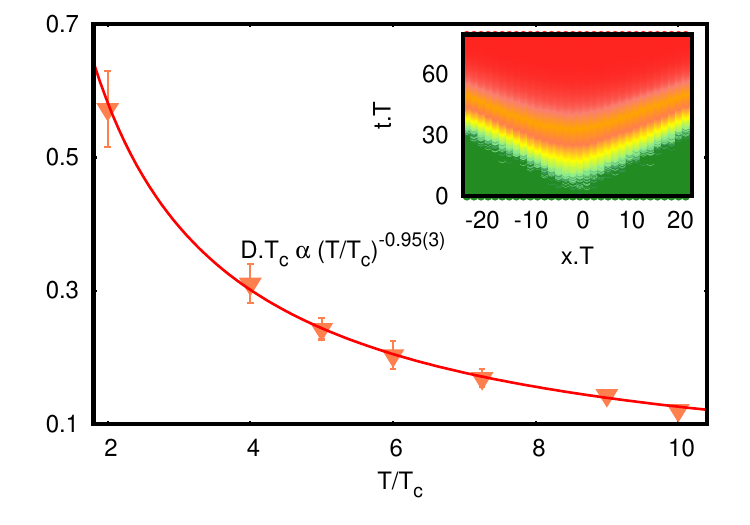}
    \caption{The temperature dependence of the diffusion coefficient $D$ in SU(2) gauge 
    theory at high temperatures $T\gg T_c$. The inset shows the ballistic spread of the  
    Lyapunov exponents in a typical thermal state of SU(2) at $T=1.2$ GeV.  The color 
    gradient from green to red denotes increasing values of the distance function in the $x$-$t$ plane.}
    \label{fig:diffusion_SU2}
\end{figure}

The diffusion coefficient $D$ which measures how fast the correlation between the fields increases with time,  
has many interesting features as a function of temperature. The $\lambda_L$ at $T_c$ shown in 
Fig.~\ref{fig:LyaScalar} correspond to the long-wavelength critical modes of SU(2) gauge theory which are 
a subset of the infrared magnetic modes, whose momenta are bounded by $|\mathbf{p}|\leq g^2T/\pi$. The fact 
that $D$ is independent of temperature below and at $T_c$ and decreases sharply above $T_c$ implies that 
these critical gluon modes at the phase transition are maximally spread in the configuration space. This 
spread is an inherently chaotic phenomena, which also manifests itself in the stochastic 
properties~\cite{Blaizot:2008nc} of the eigenvalue density of Wilson loop operator in the limit of 
large number of colors. In this limit, the quantum fluctuations are sub-dominant, and an order-disorder 
transition~\cite{Durhuus:1980nb,Narayanan:2007dv} in terms of the area of Wilson loops has similar features 
as the (de)confinement transition. 
 
The temperature dependence of $D$ at $T\gg T_c$ as shown in Fig.~\ref{fig:diffusion_SU2} can be 
interpreted from the fact that $D$ represents a diffusion in the configuration space of color 
gauge links and hence can be related to the inverse of color conductivity $\sigma$. The color 
conductivity is related to the color diffusion coefficient through the relation, 
$\sigma=\chi D$~\cite{selikhov1993color,arnold2000transport}, 
where $\chi$ is the color charge susceptibility. The $\sigma$ calculated using perturbative 
techniques in a thermal non-Abelian plasma, increases linearly with temperature~\cite{Arnold:1999uy} 
and $\chi$ varies as $\sim T^2$. Hence the diffusion coefficient varies as $D\sim \frac{1}{T}$, 
which parametrically depends on the inverse of color conductivity $\sigma$.
In the effective dynamics of the soft (magnetic) modes described by Eq.~\ref{eqn:langevinevogauge}, $\sigma$ represent 
how these modes rearrange in response to the ultra-relativistic hard modes. The soft modes receive random kicks 
from the thermal bath consisting of hard modes and a finite $\sigma$ effectively dampens them, allowing for 
efficient thermalization.

Next we address the implications of our study for understanding the mechanism of thermalization in gauge theories. 
When we discuss about thermalization, to be more precise, we imply how fast the magnetic (soft) modes equilibrate. 
Starting from a non-equilibrium state the entropy increases until it reaches its maximum in a thermal state. The KS 
entropy density rate $\Dot{s}_{KS}$ can thus be used to estimate the time required in this process. From the sum of all 
the positive Lyapunov exponents $\lambda_i$~\cite{pesin2020ljapunov}, one can estimate this quantity on the lattice as
\begin{equation}
    \Dot{s}_{KS}a^4=-\frac{\sum_{i}\lambda_i a}{N^3}\sim -4\lambda_L.{a}~,
    \label{eqn:entropy_rate}
\end{equation}
for the range of temperatures $2$-$10 ~T_c$, considering that about $1/3$ of $18N^3$ number of the Lyapunov exponents 
are positive~\cite{gong1994lyapunov} and their values can be obtained from the relation $\lambda(v)\sim \lambda_{L}[1-(\mathrm{v}/\mathrm{v}_B)^{2}]$ evident in our data, where we have taken the values of $\mathrm{v}$ as randomly distributed between $0$ and $\mathrm{v}_B$. 
We have also verified that our estimate does not vary significantly if the values of $\mathrm{v}$ are chosen to be uniformly 
distributed between $[0,\mathrm{v}_B]$. Starting from a thermal state, $\Dot{s}_{KS}$ measures the rate at which information 
about the initial state of the system is lost. We use the thermodynamic relation $s_{\text{th}} T=\varepsilon+p$ to obtain the 
entropy density of a thermal state of SU(2) to be $s_{\text{th}}a^3= \frac{4}{3}\frac{\varepsilon.a^4}{T.a}$ at 
temperatures $T\gg T_c$ where the energy density and pressure are related by $\varepsilon\simeq 3p$. On the other hand 
the entropy density for the non-thermal state in the self-similar regime described in section~\ref{sec:SU2noneq}, 
$s_{\text{non-th}}a^3=-\int{\tilde f(\vert \mathbf p\vert)\ln \left[\tilde f(\vert \mathbf p\vert)\right]~d^3 (a\mathbf p)}\simeq 4.8$ can be calculated by performing a numerical integration using the non-thermal distribution $f$ obtained 
in our computation.  
Using the entropy densities of the thermal and non-thermal states one can derive the thermalization time 
$t_{\text{th}}=0.8/\lambda_{L}(T)$, by integrating over Eq.~\ref{eqn:entropy_rate}. Thus starting from a 
highly occupied gluon state in a non-thermal scaling regime, a thermal state at $T \simeq 600$ MeV can be 
reached within a time, $t_{\text{th}}\sim 0.50(3)$ fm/c, whereas it would take $t_{\text{th}}\sim 0.70(5)$ fm/c 
to reach a thermal state at $T\simeq 450$ MeV.

\section{Outlook}

In this letter, we provide a detailed understanding of the dynamical properties of  soft (magnetic) modes of 
non-Abelian SU(2) gauge theory as a function of temperature as well as in a particular non-thermal attractor 
state. These magnetic modes interact non-perturbatively even at asymptotically high temperatures~\cite{Bala:2025ilf} 
and influence dynamical properties like the sphaleron transition rate~\cite{Bodeker:1998hm,Moore:1998zk}, 
and other transport properties of the non-Abelian plasma~\cite{Meyer:2011gj}. We show here that these modes 
play an important role in the process of thermalization in gauge theories as well. In a high temperature SU(2) plasma,
where a clean separation between the hard (ultraviolet momentum) modes and these soft modes is possible, resulting 
in an effective stochastic description of the latter~\cite{Bodeker:1998hm}, we demonstrate the chaotic behavior 
inherent in the system by extracting the spectra of positive Lyapunov exponents. This allows us to measure the 
Kolmogorov-Sinai entropy rate which determines how fast the entropy flows into the system eventually 
attaining a maximum. We have used this rate to measure a thermalization time $\sim 0.7$-$0.5$ fm/c to reach to a 
thermal state at $450$-$600$ MeV starting from a particular non-thermal state of soft gluons which shows a 
particular self-similar behavior~\cite{Berges:2013fga}. The only assumption that goes into our calculation 
is the conservation of energy density of the soft modes during the entire evolution of the system. 
Interestingly our estimate of the thermalization time in this simple non-expanding system of overoccupied 
soft gluon modes is consistent with the early switch-on time for hydrodynamics necessary to describe flow observables 
in a heavy-ion collision event~\cite{Heinz:2004pj} which is described by a glasma initial condition in an expanding 
spacetime geometry. However our result for the thermalization time is smaller than typical estimates ($\gtrsim 2.5$ fm/c) 
based on perturbative scattering process~\cite{Baier:2002bt}.

In our approach we can only predict the time taken by the soft modes to thermalize if the initial non-thermal 
and the final thermal states are known. However understanding the microscopic mechanism that drives the system 
away from the self-similar non-thermal scaling regime would require a more detailed understanding of how hard 
particles are created in this regime and how their interactions with the soft modes can be formulated within an 
effective theory. Possible mechanisms include the onset of plasma instabilities 
~\cite{Baier:2000sb,Arnold:2003rq,Romatschke:2003ms,Rebhan:2004ur,Arnold:2004ti,Kurkela:2011ub,Berges:2013eia,Berges:2013fga,Schlichting:2022fjc}, and the process of thermalization is an inherently non-perturbative phenomenon~\cite{Kovchegov:2005ss}. 
We would like to address these topics in a future work.

\section{Acknowledgements}
We are grateful to Sumilan Banerjee, Rajiv Gavai, S\"{o}ren Schlichting and Ravi Shanker for discussions related to this work. We also thank the anonymous Referee for several suggested improvements. The authors acknowledge the computing time provided by the High Performance Computing Center at the Institute of Mathematical Sciences.

\bibliographystyle{elsarticle-num} 
\bibliography{otoc_paper_v2}

\end{document}